\documentstyle[12pt]{article}

\catcode`\@=11
\def\@citex[#1]#2{%
\if@filesw \immediate \write \@auxout {\string \citation {#2}}\fi
\@tempcntb\m@ne \let\@h@ld\relax \def\@citea{}%
\@cite{%
  \@for \@citeb:=#2\do {%
    \@ifundefined {b@\@citeb}%
      {\@h@ld\@citea\@tempcntb\m@ne{\bf ?}%
      \@warning {Citation `\@citeb ' on page \thepage \space undefined}}%
      {\@tempcnta\@tempcntb \advance\@tempcnta\@ne%
      \@tempcntb\number\csname b@\@citeb \endcsname \relax%
      \ifnum\@tempcnta=\@tempcntb 
        \ifx\@h@ld\relax%
          \edef \@h@ld{\@citea\csname b@\@citeb\endcsname}%
        \else%
          \edef\@h@ld{\ifmmode{-}\else--\fi\csname b@\@citeb\endcsname}%
        \fi%
      \else
        \@h@ld\@citea\csname b@\@citeb \endcsname%
        \let\@h@ld\relax%
      \fi}%
    \def\@citea{,\penalty\@highpenalty\,}%
  }\@h@ld
}{#1}}

\def\@citeb#1#2{{[#1]\if@tempswa , #2\fi}}
\def\@citeu#1#2{{$^{#1}$\if@tempswa , #2\fi }}
\def\@citep#1#2{{#1\if@tempswa , #2\fi}}

\def\bcites{         
        \catcode`\@=11
        \let\@cite=\@citeb
        \catcode`\@=12
}

\def\upcites{         
        \catcode`\@=11
        \let\@cite=\@citeu
        \catcode`\@=12
}

\def\plaincites{      
        \catcode`\@=11
        \let\@cite=\@citep
        \catcode`\@=12
}

\newcount\hour
\newcount\minute
\newtoks\amorpm
\hour=\time\divide\hour by 60
\minute=\time{\multiply\hour by 60 \global\advance\minute by-\hour}
\edef\standardtime{{\ifnum\hour<12 \global\amorpm={am}%
        \else\global\amorpm={pm}\advance\hour by-12 \fi
        \ifnum\hour=0 \hour=12 \fi
        \number\hour:\ifnum\minute<10 0\fi\number\minute\the\amorpm}}
\edef\militarytime{\number\hour:\ifnum\minute<10 0\fi\number\minute}

\def\draftlabel#1{{\@bsphack\if@filesw {\let\thepage\relax
   \xdef\@gtempa{\write\@auxout{\string
      \newlabel{#1}{{\@currentlabel}{\thepage}}}}}\@gtempa
   \if@nobreak \ifvmode\nobreak\fi\fi\fi\@esphack}
        \gdef\@eqnlabel{#1}}
\def\@eqnabel{}
\def\@vacuum{}
\def\marginnote#1{}
\def\draftmarginnote#1{\marginpar{\raggedright\scriptsize\tt#1}}
\def\draft{
        \pagestyle{plain}
        \overfullrule=2pt
        \oddsidemargin -.5truein
        \def\@oddhead{\sl \phantom{\today\quad\militarytime} \hfil
        \smash{\Large\sl DRAFT} \hfil \today\quad\militarytime}
        \let\@evenhead\@oddhead
        \let\label=\draftlabel
        \let\marginnote=\draftmarginnote
        \def\ps@empty{\let\@mkboth\@gobbletwo
        \def\@oddfoot{\hfil \smash{\Large\sl DRAFT} \hfil}
        \let\@evenfoot\@oddhead}
        \def\@eqnnum{(\theequation)\rlap{\kern\marginparsep\tt\@eqnlabel}%
        \global\let\@eqnlabel\@vacuum}  }

\def\blackfonts{
        \font\blackboard=msbm10 scaled\magstep1
        \font\blackboards=msbm8
        \font\blackboardss=msbm6
}

\def\nblack{            
        \def\ZZ{{Z \n{10} Z}}
        \def\NN{{N \n{14} N}}
        \def\CC{{C \n{11} C}}
        \def\RR{{R \n{11} R}}
        \def\QQ{{Q \n{12} Q}}
        \def\PP{{P \n{11} P}}
}

\def\prep{         
        \catcode`\@=11
        \input art10.sty
        \catcode`\@=12
        
        \let\small\null
        \def\blackfonts{
                \font\blackboard=msbm10
                \font\blackboards=msbm7
                \font\blackboardss=msbm5
        }
        \let\sl\it
        \twocolumn
        \sloppy
        \voffset=-2.54truecm
        \hoffset=-2.54truecm
        \flushbottom
        \parindent 1em
        \leftmargini 2em
        \leftmarginv .5em
        \leftmarginvi .5em
        \marginparwidth 48pt
        \marginparsep 10pt
        \setlength{\columnsep}{2truecm}
        \setlength{\textwidth}{25.4truecm}
        \setlength{\textheight}{17truecm}
        \baselineskip=16pt
        \oddsidemargin .18truein
        \evensidemargin .17truein
}

\def\eqalign#1{\null\,\vcenter{\openup\jot\m@th
  \ialign{\strut\hfil$\displaystyle{##}$&$\displaystyle{{}##}$\hfil
      \crcr#1\crcr}}\,}
\def\eqalignno#1{\displ@y \tabskip\centering
  \halign to\displaywidth{\hfil$\@lign\displaystyle{##}$\tabskip\z@skip
    &$\@lign\displaystyle{{}##}$\hfil\tabskip\centering
    &\llap{$\@lign##$}\tabskip\z@skip\crcr
    #1\crcr}}

\def\section{\@startsection {section}{1}{\z@}{3.ex plus 1ex minus
 .2ex}{2.ex plus .2ex}{\large\bf}}
\def\subsection{\@startsection{subsection}{2}{\z@}{2.75ex plus 1ex minus
 .2ex}{1.5ex plus .2ex}{\bf}}

\def\appendix{{\newpage\section*{Appendices}}\let\appendix\section%
        {\setcounter{section}{0}
        \gdef\thesection{\Alph{section}}}\section}

\def\abstract{\if@twocolumn
\section*{Abstract}
\else 
\begin{center}
{\bf Abstract\vspace{-.5em}\vspace{0pt}}
\end{center}
\quotation
\fi}

\catcode`\@=12

\def\noj#1,#2,{{\bf #1} (19#2)\ }
\def\jou#1,#2,#3,{{\sl #1\/ }{\bf #2} (19#3)\ }
\def\ann#1,#2,{{\sl Ann.\ Physics\/ }{\bf #1} (19#2)\ }
\def\cmp#1,#2,{{\sl Comm.\ Math.\ Phys.\/ }{\bf #1} (19#2)\ }
\def\cq#1,#2,{{\sl Class.\ Quantum Grav.\/ }{\bf #1} (19#2)\ }
\def\cqg#1,#2,{{\sl Class.\ Quantum Grav.\/ }{\bf #1} (19#2)\ }
\def\ijmp#1,#2,{{\sl Int.\ J.\ Mod.\ Phys.\/ }{\bf A#1} (19#2)\ }
\def\jmp#1,#2,{{\sl J.\ Math.\ Phys.\/ }{\bf #1} (19#2)\ }
\def\grg#1,#2,{{\sl Gen.\ Rel.\ Grav.\/ }{\bf #1} (19#2)\ }
\def\mpl#1,#2,{{\sl Mod.\ Phys.\ Lett.\/ }{\bf A#1} (19#2)\ }
\def\nc#1,#2,{{\sl Nuovo Cim.\/ }{\bf #1} (19#2)\ }
\def\np#1,#2,{{\sl Nucl.\ Phys.\/ }{\bf B#1} (19#2)\ }
\def\pl#1,#2,{{\sl Phys.\ Lett.\/ }{\bf #1B} (19#2)\ }
\def\pla#1,#2,{{\sl Phys.\ Lett.\/ }{\bf #1A} (19#2)\ }
\def\pr#1,#2,{{\sl Phys.\ Rev.\/ }{\bf #1} (19#2)\ }
\def\prd#1,#2,{{\sl Phys.\ Rev.\/ }{\bf D#1} (19#2)\ }
\def\prl#1,#2,{{\sl Phys.\ Rev.\ Lett.\/ }{\bf #1} (19#2)\ }
\def\prp#1,#2,{{\sl Phys.\ Rept.\/ }{\bf #1C} (19#2)\ }
\def\ptp#1,#2,{{\sl Prog.\ Theor.\ Phys.\/ }{\bf #1} (19#2)\ }
\def\ptpsup#1,#2,{{\sl Prog.\ Theor.\ Phys.\/ Suppl.\/ }{\bf #1} (19#2)\ }
\def\rmp#1,#2,{{\sl Rev.\ Mod.\ Phys.\/ }{\bf #1} (19#2)\ }
\def\yadfiz#1,#2,#3[#4,#5]{{\sl Yad.\ Fiz.\/ }{\bf #1} (19#2) #3%
\ [{\sl Sov.\ J.\ Nucl.\ Phys.\/ }{\bf #4} (19#2) #5]}
\def\zh#1,#2,#3[#4,#5]{{\sl Zh.\ Exp.\ Theor.\ Fiz.\/ }{\bf #1} (19#2) #3%
\ [{\sl Sov.\ Phys.\ JETP\/ }{\bf #4} (19#2) #5]}

\def\beq{\begin{equation}}
\def\eeq{\end{equation}}
\def\beqar{\begin{eqnarray}}
\def\eeqar{\end{eqnarray}}

\def\nfrac#1#2{{\displaystyle{\vphantom1\smash{\lower.5ex\hbox{\small$#1$}}%
        \over\vphantom1\smash{\raise.25ex\hbox{\small$#2$}}}}}

\def\p#1{\mskip#1mu}
\def\n#1{\mskip-#1mu}

\def\comma{\p6,}

\def\lae{\mathrel{\mathop{\smash{\lower .5 ex \hbox{$\stackrel<\sim$}}}}}
\def\lae{\mathrel{\mathop{\smash{\lower .5 ex \hbox{$\stackrel>\sim$}}}}}

\def\l:{\mathopen{:}\,}
\def\r:{\,\mathclose{:}}

\def\[{\left[}          \def\]{\right]}
\def\({\left(}          \def\){\right)}
\def\<{\left<}          \def\>{\right>}

\catcode`\@=11
\def\theequation{\arabic{equation}}
\catcode`\@=12

\nblack
\bcites

\nblack

\catcode`\@=11
\def\theequation{\thesection.\arabic{equation}}
\@addtoreset{equation}{section}
\@addtoreset{footnote}{section}
\@addtoreset{footnote}{subsection}
\catcode`\@=12

\newcommand{\beqn}{\begin{equation}}
\newcommand{\eeqn}{\end{equation}}
\newcommand{\beqnarray}{\begin{eqnarray}}
\newcommand{\eeqnarray}{\end{eqnarray}}
\begin{document}

\begin{titlepage}

\begin{center}
\hfill	IASSNS-HEP-96/85 \\
\hfill	ITP-SB-96-38 \\
\hfill	hep-th/9608105

\vskip 1 cm
{\large \bf Three Dimensional Gauge Theories and Monopoles \\}
\vskip 0.1 cm
\vskip 0.5 cm
{Gordon Chalmers}
\vskip 0.2cm
{\sl
chalmers@insti.physics.sunysb.edu \\
Institute for Theoretical Physics \\ 
State University of New York, Stony Brook, NY 11794-3840 \\
}
\vskip 0.5 cm
{Amihay Hanany}
\vskip 0.2cm
{\sl
hanany@sns.ias.edu \\
School of Natural Sciences \\
Institute for Advanced Study \\
Olden Lane, Princeton, NJ 08540, USA
}

\end{center}

\vskip 0.5 cm
\begin{abstract} 
The coulomb branch of $N=4$ supersymmetric Yang-Mills gauge 
theories in $d=2+1$ is studied.  A direct connection between gauge 
theories and monopole moduli spaces is presented.  It is proposed 
that the hyper-K\"ahler metric of supersymmetric $N=4$ $SU(N)$
Yang-Mills theory is given by the charge $N$ centered moduli space of BPS 
monopoles in $SU(2)$.  The theory is compared to $N=2$ supersymmetric
Yang-Mills theory in four dimensions through compactification on
a circle of the latter.  It is found that rational maps are
appropriate to this comparison.  A BPS mass formula is also written
for particles in three dimensions and strings in four dimensions.
\end{abstract}

\end{titlepage}

\section{Introduction}
Recently there has been some interest in the dynamics of 
gauge theories in three dimensions. In \cite{nati} the dynamics of 
$U(1)$ and $SU(2)$ gauge theories was
derived from string theory by looking at compactification of 
type I theory on $T^3$.  This work was motivated by an 
interesting idea that branes can probe the geometry
of space time \cite{douglas} and was applied to four dimensions 
in \cite{bds} following the work of \cite{sen} on F-theory and 
orientifolds. In \cite{ed} a detailed discussion of the dynamics 
of $U(1)$ and $SU(2)$ gauge theories and comparison to expectations 
from string theory was performed. It was found that the Coulomb branch 
of the moduli space for low number of flavors is given by the moduli
space of two monopoles. The aim of this paper is to generalize some of the
results to $SU(N)$ Yang-Mills theories. We will make a direct connection
with the moduli space of charge $N$ $SU(2)$ monopoles.

In a recent work \cite{hk} we looked at an example of two 
intersecting five branes in M-theory which intersect over a 
3-brane. The theory on the intersection is associated with the 
appearance of tensionless strings in four dimensions.  The theory
has two phases which are connected by the tensionless string 
point. One in which the branes are on top of each other and it 
was argued in \cite{hk} that they form a bound state. Further 
interpretation of this phase is still lacking. It was also
argued that at the point of intersection a $U(1)$ gauge field 
and a hypermultiplet are unHiggsed and thus possibly the theory 
on the intersection has a Coulomb phase corresponding to the
$U(1)$ field. The other phase is when the two branes are 
separated. This phase has a moduli space of vacua which is 
parameterized by a linear (tensor) multiplet in four dimensions.  
Further reduction of the two five brane configuration to two 
four branes intersecting over a two brane leads to the study
of a gauge theory in three dimensions. There appear to be many 
similarities between the linear multiplet moduli space in 
four dimensions and the Coulomb branch of the
three dimensional $U(1)$ gauge theory. This similarity has presumably 
origins in string 
theory. In particular the moduli space of those theories looks like 
a two monopole moduli space. So a natural question is to ask do we see
the monopoles in the string theory? A partial answer can be given by
viewing the two five branes which lie in say 12345 and 12367 directions 
as two magnetic monopoles which move in the 8,9,10
directions. Compactification of the 3-rd direction which leads from the
four dimensional theory to the three dimensional theory then provides the
reason why the two moduli spaces are similar. The above observation also
suggests that the Coulomb branch of $N=4$ gauge theories in three
dimensions is connected to the moduli space of $n$ monopoles where $n$ is
related to the rank of the gauge group.

The aim of this paper is to study the relation of $SU(N_c)$ gauge 
theories in three dimensions to a system of $N_c$ monopoles. In 
particular we find that semiclassically the Coulomb branch of $SU(N_c)$ 
gauge theory is the moduli space of $N_c$ monopoles when the monopoles 
have large separations. Moreover we identify the moduli space of $SU(N)$
Yang-Mills with the moduli space of $N$ monopoles.

The outline of the paper goes as follows.
In section 2 we describe the three dimensional models and
calculate the semiclassical metric.  We also relate them to 
gauge theories on $\RR^3\times S^1$ with radius $R$.
We discuss the structure of the moduli spaces and make the 
comparison to the moduli space of $SU(2)$ monopoles.  In section 3 
we discuss the $R$-dependence of the metric.  We derive a
formula for large $R$ which relates the period matrix (matrix of couplings
and theta angles) associated to four dimensional
gauge theories with $N=2$ supersymmetry to the Dirac connection on
the moduli space of monopoles.  In section 4 we review   
the correspondence of monopoles with the rational maps 
and discuss their application to three dimesional gauge theories. The 
picture associated with rational maps has a very
natural role in connecting the three dimensional $N=4$ gauge theories to
the four dimensional $N=2$ gauge theories. We discuss their implications.
We give a few examples and introduce a BPS mass formula in three
dimensions.  This formula is also related to a BPS tension of a string
in four dimensions.

While writing this paper two papers appeared \cite{si,ss} which
discuss problems related to the problems discussed in this paper.

\section{$N=4$ SU(N) gauge theories in three dimensions} 

Three dimensional gauge theories with $N=4$ supersymmetry have three types 
of global symmetries. It is convenient to perform a dimensional reduction 
from $N=1$ supersymmetric gauge theories in six dimensions in order to 
understand the symmetries.  There is an $SU(2)_R$ symmetry which acts 
on the six dimensional fermions.  The fermions and supercharges transform 
as doublets of this symmetry.  The vector multiplet in three dimesions 
contains three scalars $\phi_i,$ $i=1,2,3.$  There is a rotation group 
under which these scalars transform as a vector.  We will denote its 
double cover $SU(2)_N$. The last symmetry is the rotation group of three 
dimensional Euclidean space $\RR^3$. Its double cover will be 
denoted $SU(2)_E$.

The matter content of a $N=4$ super Yang-Mills theory consists of a vector
multiplet which transforms in the adjoint representation of the 
gauge group
$G$. We will be interested in this paper in studying the Coulomb branch in
the case where the gauge group is $SU(N_c)$. Under $SU(2)_R\times
SU(2)_N\times SU(2)_E$ the fermions of a vector multiplet transform as
({\bf2},{\bf2},{\bf2}) and the scalars transform as ({\bf1},{\bf3},{\bf1}).

The potential energy for the scalars is
\beq
V={1\over e^2}\sum_{i<j}Tr[\phi_i,\phi_j]^2, 
\label{potential}
\eeq
where $e$ is the gauge coupling.  The potential $V$ vanishes if 
the $\phi_i$ commute; the space of zeros of $V$ has dimension
$3r$ where $r$ is the rank of the gauge group. At a generic point in 
moduli space the gauge group is broken to $U(1)^r$. In addition to 
the $3r$ scalars we have also $r$ massless photons which are each 
superpartners of the three scalars. Those photons are dual to compact 
scalars so the moduli space of the Coulomb branch is parameterized 
by $4r$ scalars.  The low-energy effective action is a linear sigma 
model with target space a hyper-K\"ahler manifold of quaternionic 
dimension $r$. In addition to the vector multiplets we will at times couple 
the theory to $N_f$ hypermultiplets which transform in the 
fundamental representation of the gauge group.

The low-energy (leading) component of the N=2 theory in d=3+1 
dimensions is written in terms of a chiral and vector $N=1$ 
superfield $\phi^a$ and $W_\alpha^a$.  The low-energy theory 
before compactification is 
\beq 
S=\Im \int d^4x d^2\theta {1\over 2}{\cal F}_{ab}(\phi) W^{a\alpha} 
 W^b_\alpha + \Im \int d^4xd^2\theta d^2{\bar\theta} {\cal F}_a (\phi) 
 {\bar \phi}^a  \ .  
\eeq 
We only consider the fields living in the Cartan sub-algebra of the 
bosonic sector, in which the action is given by 
\beq 
S_g = \int d^4x {1\over 4e_{ij}^2} F^i_{\mu\nu} F^{j,\mu\nu} 
 + {i\theta_{ij}\over 32\pi^2} F^i_{\mu\nu} {\tilde F}^{j,\mu\nu} 
 + \int d^4x {1\over 2e_{ij}^2} 
   \partial_\mu {\bar\phi^i}\partial^\mu\phi^j   \ .
\eeq 
We have kept only terms arising in the action which 
have two derivatives; the terms in $(D_\mu\phi)^{i,\dagger} 
(D^\mu\phi)^j$ other than the kinetic one do not contribute 
to the low-energy effective action of the scalars.   

The effective three dimensional theory is found by 
compactifying on a spacetime $R^3\times S^1$ and keeping only the 
$r$ photons and the $r$ scalars arising from the fourth component 
of the $4$-dim gauge fields.  We further will denote the 
complex scalars in the chiral $N=1$ multiplet by the 
real pair $\phi^i=\phi^i_1+i\phi^i_2$. 

In the following we give the description of the moduli space 
of the three-dimensional gauge theory $SU(N_c)$ in the semi-classical 
regime.  The space of fields $\phi_i$ which minimize the scalar 
potential in eq.(\ref{potential}) can be parameterized in the Cartan 
sub-algebra as
\beq
\phi_i={\rm diag}\[x^{i}_1,...,x^{i}_{N_c}\] \,
\label{phi}
\eeq
with the condition that $\sum_{j=1}^{N_c}x_{ij}=0, i=1,2,3.$ This
parameterization is up to a Weyl transformation and so the space of 
zeroes of $V$ is given by $\RR^{3N_c}/S_{N_c}$ restricted to the above 
condition.  The classical moduli space of the scalar fields, 
however, is required not to have any points of enhanced gauge 
symmetry.  These points exist whenever the discriminant 
\beq 
\Delta = \prod_{i\neq j} \vert {\vec\phi}_i - {\vec\phi}_j \vert 
\label{disc}
\eeq
vanishes. 
In the quantum theory we expect the metric at 
these points to be smooth or singular depending on the number 
of flavors 
Semi-classically we delete the points in which at least two 
of the scalar field expectation values are the same.  There 
are ${1\over 2}N_c(N_c-1)$ surfaces $\Delta_{ij}$ 
defined by the condition on the pairs ${\vec\phi}_i= {\vec\phi}_j$ 
in which we delete.  The 
classical moduli space of a completely broken gauge theory is then 
${\cal M}_{cl} = \left(\RR^{3N_c}-\{\Delta_{ij}\}\right)/S_{N_c}$.  
$S_{N_c}$ is the permutation group for $N_c$ elements.  Under 
the action of an element $P_{ij}$ which exchanges the fields 
${\vec\phi}_i \leftrightarrow {\vec\phi}_j$ the corresponding deleted 
surfaces in $\RR^{3N_c}$ are identified.  There are points of enhanced 
gauge symmetry whenever scalar expectation values 
meet; for example, points in which $M$ scalar vevs are equal give 
rise to a symmetry group $SU(M)\times U(1)^{N_c-M}$ although 
various product groups are possible. 

The Weyl group also acts on the compact scalars dual to the photons
$\theta_i, i=1,\ldots N_c$, $\sum_{i=1}^{N_c}\theta_i=0$ by 
an exchange in $S_{N_c}$.  Including these scalars the classical 
$4N_c$-dimensional moduli space of completely broken vacua is  

\beq 
{\cal M}_N^{{\rm cl}}= {(\RR^{3N_c}-\Delta_{ij}\}) 
  \times T^{N_c} \over S_{N_c} } . 
\label{class}
\eeq 
One-loop corrections generate a non-trivial Dirac 
connection on the $T^{N_c}$ given by $d\theta_i+\Gamma_{ij}^k 
dx_k^{j}$, as described below, and the $T_{N_c}$ is promoted to 
a non-trivial fibre.  Classically this vanishes and the vacua is simply 
a direct product of the compact scalars on $T^{N_c}$ together 
with the scalars on $R^{3N_c}$.   

The classification of the possible $T^{N_c}$ fibre bundles with the 
appropriate symmetry under the permutation group is a generalization 
of the results in \cite{ed}.  The base, $\RR^{3N_c}-\Delta_{ij}$, 
in eq.(\ref{class}) has a second homology group 
$H_2({\cal M}_{N_c}^{\rm cl},Z)$ with dimension ${N_c (N_c-1)\over2}$.  
A homology basis is given by the 2-spheres $S^2_{ij}$ 
defined by $|\vec x_i-\vec x_j|=r_{ij}={\rm const}$, $x_k={\rm const}$, 
$i\not=j\not=k$.  In order to determine the fiber we must  
specify the connection one form associated to the $i$-th generator of 
$T^{N_c}$ around the different homology generators.  However, 
examining the connections restricted to each of the  
$S^2_{ij}$ together with the permutation symmetry reduces 
it to the standard Dirac monopole connection 
over $S^2$ with a magnetic charge $s$; the action of the 
permutation group implies the topological index is the same 
around each of the spheres.   

In the quantum gauge theory the integer $s$ which labels the 
non-trivial semi-classical $T^{N_c}$ fiber at infinity may be found 
either by a one loop computation or by a counting of the fermion 
zero modes which give rise to an instanton correction.  It 
is noteworthy that in three dimensions magnetic monopoles in 
an unbroken $U(1)$ subgroup of the gauge group may appear as 
instantons. It was found \cite{ed} to be $s=4$ for an $SU(2)$ group
by analyzing the symmetries of the instanton.
This analysis is independent of the gauge group.
Furthermore by counting the zero modes of a vector multiplet
in the presense of one monopole \cite{CH} it is found to be
the same for arbitrary gauge group.
\beq
s=4.
\eeq
Due to such an instanton the $SU(2)_N$ symmetry is broken down to 
$U(1)_N$ by the expectation value of the scalars.  Under this 
symmetry there are $4$ zero modes with charge ${1\over2}$ which 
come from the gauge multiplet.  An $r$-instanton 
contribution to the metric will then be possible if
\beq
4r=4 .
\eeq
This implies that $r=1$.
There can be also contributions from anti-monopoles which will 
have zero modes with opposite $U(1)_N$ charges.  The counting 
then does not change essentially, $r$ is replaced with 
the net instanton number.  Note that this counting implies 
that the value for the coefficient of the Dirac potential 
in the semi-classical result above to be $s=4$.
Indeed, the asymptotic form of the metric for a charge $k$ monopole
has a coefficient of the Dirac potential which is independent of $k$
\cite{gm}. Thus, the asymptotic forms of the monopole and gauge
theory metrics coincide and lends further support to the identification.

The classical vacuum consisting of the $3N_c$ scalars and $N_c$ 
compact duals to the photon has a flat metric described by 
\beq 
ds^2=\sum_{i=1}^{N_c}({1\over e^2}d\vec x_i^2+e^2d\theta_i^2) . 
\eeq   
where the Weyl group action is implied.
As was found in \cite{nati} one loop corrections change 
the metric in the asymptotic regime.  Considering a $U(1)$ gauge group 
the one-loop effect gives rise to a non-trivial connection 
on the periodic scalar in $T$ and is
\beq 
d\theta\rightarrow d\theta-s\vec w\cdot d\vec x. 
\eeq 
The vector $\vec w(\vec x)$ satisfies the field equation of a 
Dirac point monopole 
\beq 
\nabla\times\vec w=\nabla({1\over|\vec x|}) . 
\eeq 
$N=4$ supersymmetry requires the metric to be hyper-K\"ahler.  Necessary
conditions for the metric to be hyper-K\"ahler \footnote{see \cite{gm} and
references therein.} then lead to the corresponding correction to the 
metric on $R^3$ given by 
\beq 
{1\over e^2}\rightarrow{1\over e^2}-{s\over|\vec x|}. 
\label{correct}
\eeq
The metric is semi-classically corrected to be of the Taub-Nut type 
with a negative mass parameter for positive $s$.  It possesses 
a real singularity, i.e. it is not geodesically complete, when the 
latter side of equation (\ref{correct}) is zero; this occurs 
for positive $s$.  

Note that in the limit of large fields the semi-classical correction 
may be neglected and the metric reduces to the flat one given in 
\cite{nati}.  The total correction can also be computed by a one loop 
calculation as it has power of $e^0$.  Dimensional reasoning leads 
to the form of this correction and the coefficient $s$ is the 
analogue of the beta function in four dimensions (it is proportional 
to the corresponding zero mode counting in $d=4$).  

We now generalize the semi-classical results to other examples, 
including $SU(N_c)$.  We begin with the $U(1)$ gauge theory 
$s=-N_f,$ where $N_f$ is the number of hypermultiplets \cite{ed}.
For $U(2)$ the asymptotic metric has the form
\beq
ds^2=Gd\vec x_-^2+d\vec x_+^2+G^{-1}d\tilde\theta_-^2+d\theta_+^2, 
  \qquad  G={1\over e^2}-{s\over r_{12}},
\label{utwo}
\eeq
with the following notations
\beq 
\vec x_\pm={1\over\sqrt2}(\vec x_1\pm\vec x_2),\qquad
\theta_\pm={1\over\sqrt2}(\theta_1\pm\theta_2), 
\eeq 
\beq 
r_{12}=|\vec x_1-\vec x_2|,\qquad \vec w_{12}=\vec w(\vec x_1-\vec x_2), 
\eeq 
\beq 
d\tilde\theta_-=d\theta_--s\vec w_{12}\cdot d\vec x_-. 
\eeq  
The coordinates with subscript $+$ represent the $U(1)$ fields which 
decouple from the $SU(2)$ part.  For a configuration of two monopoles 
this corresponds to a decoupling of the center of mass motion.  It 
has been proven in \cite{ed} that there are no further instanton 
corrections to the semi-classical results.   

We generalize the semi-classical results for $SU(N_c)$ groups, with a 
$4(N_c-1)$ dimensional moduli space, by generalizing the one-loop 
calculation and writing equation (\ref{utwo}) in the form 
\beq
ds^2=g_{ij}d\vec x_i\cdot d\vec x_j+(g^{-1})_{ij}d\tilde\theta_id 
 \tilde\theta_j,
\label{un}
\eeq
\beq
d\tilde\theta_i=d\theta_i-{s\over2} 
  \sum_{j=1}^{N_c}\vec W_{ij}\cdot d\vec x_j . 
\label{thetan}
\eeq 
The function ${\vec W}_{ij}$ is the potential arising from a 
Dirac point monopole at $i$ at the point $j$.  As such, we 
have a non-trivial connection on the $T^{N_c}$ fibers given 
by
\beq 
W_{ii}=-\sum_{i\not=j}w_{ij}, 
\qquad 
W_{ij}=w_{ij},\qquad i\not=j.
\label{asymmet}
\eeq 
Given the form of ${\vec W}_{ij}$, the hyper-K\"ahler condition 
for the metric in equation (\ref{un}) as discussed in
equation ({\ref{hyper}) 
enforces the remaining components of the metric to change to 
\beq 
g_{ii}={1\over e^2}-{s\over2}\sum_{i\not=j}{1\over r_{ij}}
\qquad 
g_{ij}={s\over2}{1\over r_{ij}},\qquad i\not=j . 
\label{nondiag}
\eeq
The metric (\ref{utwo}) for $U(2)$ now takes the form given by equations
(\ref{un}), (\ref{thetan}) and
\beq
g=\pmatrix{{1\over e^2}-{s\over2r_{12}} & {s\over2r_{12}}\cr
{s\over2r_{12}} & {1\over e^2}-{s\over2r_{12}} \cr},\qquad \vec W=\vec
w_{12}\pmatrix{-1 & 1\cr 1 & -1 \cr}. 
\label{utwoo}
\eeq  
Viewing each pair of distinct indices $i,j$ as defining an $SU(2)$
subgroup we can replace the indices $\{1,2\}$ with $\{i,j\}$ and get 
the form of equations (\ref{nondiag}) verifying 
the metric for $SU(N_c)$.

The three K\"ahler forms $\omega^a$ associated with the above 
semi-classical metrics have a simple explicit description: 
\beq 
\omega^a=-{1\over 2} g_{ij} \epsilon^{abc} dx_i^b \wedge dx_j^c 
 + (d\theta_i - {s\over 2} {\vec W}_{ij}\cdot d{\vec x}_j^a) 
  \wedge dx_i^a \ . 
\label{forms}
\eeq 
Under the $SO(3)$ rotational group these forms transform as 
a vector.  Furthermore, in the absence of instanton corrections 
(plausibly in cases $N_f>N_c$ in $SU(N_c)$) the K\"ahler forms in 
eq.(\ref{forms}) are exact in the quantum theory. 

The metrics in eq.(\ref{un}) are special in that they 
possess $N_c$ isometries generated 
by constant translations of the $N_c$ coordinates $\theta_i$.  
These isometries preserve all three K\"ahler forms $\omega_i$ of the 
hyperk\"ahler metric (i.e. tri-holomorphic).  Physically they correspond 
to the independent conservation of the $U(1)^N$ factors living in the 
Cartan sub-algebra of the gauge group.  
(Examples of the complete forms of these metrics have recently been
used to describe moduli spaces of distinct fundamental monopoles in
higher-rank gauge groups \cite{Weinberg}).
Furthermore, a $4N$ dimensional 
metric with the maximal number of tri-holomorphic isometries, i.e. 
$N$, may always be written in the form in eqs.(\ref{un}) and 
(\ref{thetan}) with 
\beq 
{s\over 2}\bigl( 
\partial_i^{(a)} W_{jk}^b - \partial_j^{(b)} W_{ik}^{(a)} \bigr) 
= \epsilon^{abc} \partial_i^{(c)} g_{jk} 
\qquad 
\partial_i^{(a)} g_{jk} = \partial_j^{(a)} g_{ik} \ , 
\label{hyper}
\eeq 
where $\partial_i^{(a)}\equiv \partial/\partial x^a_i$ \cite{LR}.  
One may 
verify that the solution to the metric above satisfies these 
conditions.  We do not expect this relation to hold for $N_f<N_c$ flavors, 
as we expect non-perturbative instanton corrections to generate 
mass terms for the $U(1)$ gauge fields \cite{AM}, thus 
breaking the isometries.  

It is also useful as a consistency check to examine the limit to 
partial symmetry breaking of the 
semi-classical results; in the solution (\ref{un}) we can test the
breaking of the $SU(N_c)$ theory down to $SU(N_c-1)\times U(1)$ by 
setting $N_c-1$ of 
the fields to have equal expectation values. Explicitly we set
\beq
\vec x_i=\vec x+\vec y_i,\quad i=1,\ldots,N_c-1, 
 \qquad\vec x_{N_c}=(1-N_c)\vec x,
\eeq
where $y_i$ are very small compared to $\vec x$ and satisfy
$\sum_{i=1}^{N_c-1}\vec y_i=0$.  The metric then has terms which 
for large $x$ goes to zero as ${1\over |\vec x|}$. Neglecting this 
term produces the expected result that the $x_{N_c}$ field decouples 
to a $U(1)$ field and the rest form the metric for the $SU(N_c-1)$ 
theory.  At an enhanced $SU(2)$ subgroup, namely when two 
coordinates $x_i,x_j$ coincide while the other are far away, we 
expect to get the structure as in equation (\ref{utwoo}), which is 
also verified.

Note that for positive $s$ and sufficiently small fields the 
metric is no longer positive definite.  The semiclassical 
approximation breaks down in this regime.  However, it is known 
that there are further contributions to the metric which are 
exponentially small when the fields are large; they arise from 
monopole/instanton corrections in $d=2+1$ and cure this 
problem which is seen asymptotically.  

\vskip .2in
\noindent {\it Monopole Comparison}

The low-energy dynamics of a charge $k$ monopole in an 
$SU(2)$ theory broken down to $U(1)$ is modeled by geodesic 
motion on a hyperK\"ahler manifold of dimension $4k$.  The 
metric factorizes for general $k$ into the structure 
\beq 
M_k = {{\tilde M}^0_k \times S^1 \over Z_k} \times \RR^3 
\eeq 
The $\RR^3$ and the $S^1$ are related to the physical 
decoupling of the overall center of mass and phase of the 
$k$-monopole.  The $k$-fold cover of the moduli space 
breaks isometrically into a direct product of the form 
${\tilde M}_k = {\tilde M}^0_k \times S^1 \times \RR^3$.  In 
the asymptotic regime of moduli space where the $k$ monopole 
may be thought of as separated charge one solitons, the 
$k$-fold cover of the metric asymptotically approaches 
the form 
\beq 
{\tilde M}_k \rightarrow (\RR^3\times S^1)^k 
\eeq 
in which the parameters roughly label the positions and 
(asymptotically independent) $U(1)$ phases of the $k$ monopoles 
possessing magnetic charge one.  Furthermore, 
the reduced (or centered) moduli space of a charge $k$ 
monopole is of quaternionic dimension $k-1$ and is 
denoted by $M_k^0$; the $k$-fold cover is ${\tilde M}_k^0$.  

On $M_k$ the translations in $\RR^3$ and rotations of $SO(3)$ 
act as isometries.  Furthermore, the $SO(3)$ action rotates the three 
complex structures $\omega_i$ on the hyperk\"ahler manifold 
$M_k$ into eachother.  We briefly discuss the $Z_k$ action.  The 
periodic $S^1$ coordinate $\psi$ is associated with the 
unbroken $U(1)$ of the 
monopole in the sense of being a conjugate ``momentum'' 
coordinate to the electric charge.  The $Z_k$ group acts 
on this coordinate by $\psi\rightarrow\psi+{2\pi\over k}$ together 
with a non-trivial action on the covering ${\tilde M}_k^0$.  
For the four-dimensional centered moduli space, $M_2^0$, the 
constraint of having an $SO(3)$ group of isometries which 
rotates the complex structures constrains the form to be uniquely 
determined.   
 
The one loop corrected metric (\ref{asymmet}) coincides with the
asymptotic metric for $N_c$ monopoles which are located far apart from
each other \cite{gm}. The three real scalar expectation values of the
gauge theory correspond to the asymptotic positions of the monopoles.
The scalar dual to the photon is identified with the large gauge
transformation corresponding to each of the monopoles.  The 
asymptotic topology of the moduli space $M_k$ is 
of the same structure as the corresponding gauge theory in 
three dimensions.  This correspondence goes further and we discuss 
it in the next sections.

\section{R-Dependence} 

In this section we discuss the gauge theory on a four dimensional manifold
given by $\RR^3\times S^1$ where the $S^1$ has radius $R$.
We expect to find the known $N=2$ $d=4$ results in the limit
$R\rightarrow\infty$ while the $R\rightarrow0$ limit should produce the
three dimensional gauge theories discussed in this paper.
The scalars in the theory are the fourth components of the gauge fields,
the dual photons, and the complex superpartners of the gauge fields.
For a gauge group with rank $r$ they parameterize locally $T^{2r}\times
\CC^r$. 

The $SO(3)_N$ symmetry which acts on the scalars in the $R=0$
limit is broken down to $SO(2)_N$ for $R\rightarrow\infty$, 
which is usually called $U(1)_R$ in $d=4$.  There are a $S^2$ 
worth of different ways in comparing the compactifications 
of the $N=1$ theory in six dimensions down to $d=3$ and $d=4$.  
We dimensionally reduce on the coordinates $x^4,x^5,x^6$ in 
compatifying to the $N=4$ three-dimensional theory, but in arriving 
to the $N=2$ four-dimensional result one may dimensionally reduce 
on any plane orthogonal to a unit vector in $S^2$.  There is a 
natural correspondence between rational maps and monopoles which 
involves choosing a preferred direction.  For this reason it is 
very convenient to look at rational maps in order to study the 
$R$ dependence of the metric on the moduli space and also
to relate the two extreme limits of zero and infinite radius.
We will introduce the rational maps in the next section and in this
section we discuss the $R$-dependence of the theories.

Performing a duality transformation 
on the photons similar to that in \cite{ed} we obtain the 
low-energy theory on $\RR^3\times S^1$ 
\beq 
L= {1\over \pi R e^2_{ij}} db_i db_j + {e^2_{ij} \over \pi R (8\pi)^2} 
 d{\tilde\sigma}^i d{\tilde\sigma}^j  + {\pi R\over e_{ij}^2} 
 \partial_\mu {\bar\phi}^i \partial^\mu \phi^j 
\eeq  
with 
\beq 
d{\tilde\sigma}^i = d\sigma^i - {\theta^{ij}\over \pi}db_j 
\label{classconn} \ .
\eeq
We can now compare the form of the metric induced by this
action to the form of the semi-classical metric calculated in section 2.

As was explained in \cite{ed}, there exist a distinguished complex 
structure $\omega$ in which there is no $R$ dependence.  This essentially 
follows from the holomorphy of the complex structure.  We associate 
this particular combination of the three complex structures 
in the $N=4$ model (at $R=0$) with the choice of $x^4$ being the  
direction of the compactification on $S^1$, which corresponds to the 
third direction of the vector multiplet of $SO(3)$ of scalars 
in the notation of section two.  We identify this prefered 
complex structure for all $R$ as arising from the three-dimensional 
theory with the natural one arising on the rational maps defined 
along the preferred direction.  This allows us to identify the fields
$b_i$ with the fields in the third direction $x^3_i$. It also allows us
to identify the matrix of theta angles $\theta_{ij}$ as the Dirac
connection in the third direction $W^3_{ij}$ and the matrix of coupling
constants $e^{-2}_{ij}$ as the metric on the moduli space $g_{ij}$.

In the theories that do not possess any instanton corrections 
for any $R$, we can use the hyperK\"ahler constraints, equation
(\ref{hyper}) to derive a condition on the Dirac connection of the theory
to the period matrix of couplings and theta angles. We set
$W_{ij}=W^1_{ij}-iW^2_{ij}$ and $z_j=x^1_j+ix^2_j$ and the constraint
takes the form
\beq
\partial_{z_i}\tau_{jk}=\partial_j^3W_{ik}
\eeq 
This relation is an exact statement for these theories between 
$R=0$ and $R=\infty$.  

In cases where instantons correct the metric there is a
breaking of the $U(1)$ symmetries and the semi-classical form of 
the metric on the moduli space, given in section two, no longer 
holds. It would be interesting to derive a similar
relation for such cases, although in general the metrics possessing 
instanton corrections do not have the appropriate $U(1)$ isometries 
to be obtained by the condition in eq.(\ref{hyper}).

In the dualized theory the structure of the classical moduli 
space is given by a $N$ fibration over $C^r$ with an $R$ 
dependence.  This is seen in the semi-classical form of the metric in 
eq.(\ref{classconn}).  In the case of $SU(2)$, for example, 
the structure of the quantum vacua is an elliptic fibration 
over $C^2$ with the area of the torus fiber $V_E=1/16\pi R$ 
(in units where $\tau={\theta\over\pi} + {8\pi\over e^2}$) and 
in the normalization of eq.(\ref{classconn}).  
Semi-classically, the periodicity of the 
coordinates $b_i$ and $\sigma_i$ for all $SU(N_c)$ gauge 
groups together with the form of the metric in eq.(\ref{classconn}) 
gives the corresponding structure of $N$ as a $T^{2r}$ bundle 
over $C^{r}$.  The corresponding volume of $N$ for higher rank 
gauge groups goes accordingly as 
$V_N\sim {1/R^r}$ and diverges as $R\rightarrow 0$.

\section{Rational Maps}

We would like to consider the structure of the quantum moduli spaces 
of vacua for $N_f=0$ in $d=2+1$ through a generalization of the $SU(2)$ 
result.  There a curve was introduced describing the (non-compact) 
Atiyah-Hitchin 
manifold, on which the physical $U(1)_R$ symmetry has a simple 
action.  The analog for higher 
rank gauge groups is to consider a curve describing the non-compact 
moduli space.  There is to every monopole configuration 
an associated space, the rational mappings from $CP^1 \rightarrow 
CP^1$ together with a constraint of a determinant equation; in this 
approach of looking at the moduli space the $U(1)_R$ symmetry 
of the vacuum is easily seen and the structure of the $R>0$ of the 
compactification scale on $S^1$ is more clear.     

A based rational map in $R_k(z)$ describing a charge $k$ monopole is 
defined as
\beq 
S(z) = {p(z)\over q(z)} ={\sum_{j=0}^{k-1} a_j z^j \over z^k +
\sum_{j=0}^{k-1} b_{k-j} z^j} \ .
\label{pq}
\eeq
A theorem of Donaldson \cite{donaldson}
states that there is a one-to-one correspondence
between the universal cover of a $k$-monopole configuration (in SU(2)) 
and a map in $R_k$.  Note
that there are $k$ complex solutions $\beta_i$ to $q(z)=0$ and another $k$
values from the numerator $p(\beta_i)$.  The rational map contains the $4k$ 
real dimensional space describing the monopole configurations.  Requiring 
non-degeneracy of the $2k$ complex coordinates means that the values 
$\beta_i$ and $p(\beta_i)$ be different.  The discriminant of such a map 
defines the ``resultant, 
\beq 
\Delta_k = \prod_i p(\beta_i) \neq 0 
\eeq 
which is a non-vanishing complex number.    

It is of physical interest to explain the meaning of $R_k$.  
The rational map associated to a $k$-monopole may be found through 
the following.  We pick a direction $\vec{u}$ in $R^3$ and consider 
the equation $D_{\vec{u}} s=(\nabla_A - i\Phi)s=0$ for values in $R^3$ 
orthogonal to $\vec{u}$ (parameterized by $z\in C$).  We further take 
the parameter $t$ to label the coordinate along $\vec{u}$ and we 
use a connection for $\nabla_A$ arising from the vector potential 
along $\vec{u}$.  

We consider complex solutions to $s$.  The solution set along the 
oriented line $\vec{u}$ is two dimensional and is usually denoted 
$E_{\vec{u}}$; the total space of all solutions defines a smooth 
complex vector bundle $E$ fibred over the space of oriented lines 
in $R^3$.  
One now defines two sub-bundles $E^{\pm}$ with fibres $E^{\pm}_{\vec{u}}$ 
to be the solution space to $D_{\vec{u}}s=0$ which decay 
exponentially along $\pm\vec{u}$ at infinity, respectively.  The 
set of all $\vec{u}$ in which $E^+_{\vec{u}}=E^-_{\vec{u}}$ 
define a curve in the space of all oriented lines in $R^3$.  The 
space of oriented lines in $R^3$ is isomorphic to $TCP_1$ and the 
special oriented lines are denoted spectral lines.  The defining 
equation, the spectral curve given in equation (\ref{spectral}),
then defines a 
collection of spectral curves given by a sub-manifold in $TCP_1$.
  
The connection with the rational map is made by considering the 
scattering data associated with the solution space along the spectral 
lines.  There are two linearly independent solutions $s_0$, $s_1$ as 
$t\rightarrow\infty$ to the operator $D_{\vec{u}}s=0$ which 
have the asymptotic form given by  
\beq 
s_0(t) t^{-k/2} e^t \rightarrow c_0 \quad\quad 
s_1(t) t^{k/2} e^{-t} \rightarrow c_1  \ .
\eeq
The parameters $c_0$ and $c_1$ are constants, and we see that the 
solution $s_0$ decays exponentially.  Let ${\tilde s}_0$ be the 
corresponding solution which decays at $t\rightarrow -\infty$.  Then, 
\beq 
{\tilde s}_0 = a(z) s_0(z,t) + b(z) s_1(z,t) 
\eeq 
where $b(z)$ is a polynomial of at most degree $k$ \cite{AH}.  
The scattering information contained in $a(z)$ and $b(z)$ define 
the rational map through the relation $S(z)=p(z)/b(z)$, where 
$p(z)$ is the unique degree $k-1$ polynomial which is $a(z)$ 
modulo $b(z)$.   

The definition of the rational map and scattering data for the 
$k$-monopole required a preferred direction in $R^3$.  The identification 
between the gauge theory vacua and the monopole moduli spaces is made 
more clear if we list the transformations on the map arising from: 

\begin{enumerate} 
\item SO(2) rotations about the direction $\vec{u}$, 
\item global U(1) rotations corresponding to total charge, 
\item translations along $\vec{u}$, 
\item translations along the plane orthogonal to $\vec{u}$.  
\end{enumerate} 

\noindent The group is $G={\rm SO}(2)\otimes U(1)  
\otimes R \otimes C \sim SO(2) \otimes C^* \otimes C$, and we define 
the element by $(\lambda,\mu,\nu)$ where $\ln\vert\mu\vert \in R$ 
($\mu\neq 0$) and ${\mu\over \vert\mu\vert}\in U(1)$ and $\lambda$ a 
phase.  The total group operation induces a change in the rational 
map by 
\beq 
S_k(z) \rightarrow \mu^{-2} \lambda^{-2k} S_k({z-\nu\over\lambda}) \ . 
\eeq
Centering the $k$-monopole by fixing the center-of-mass and U(1) 
coordinate will leave us with a remaining SO(2) action. 

We write the rational map in the form where the simple poles 
are manifest
\beq 
S(z) = \sum_{i=1}^k {\alpha_i \over z-\beta_i} \,
\label{alpbet}
\eeq
In the case of well-separated monopoles it is known that 
an approximate location for each of them is given by 
$\vec{x}_i$ in $R^3$ by 
\beq 
(\Re (\beta_i) ,\Im(\beta_i) , 
-(1/2k)\ln(\vert p(\beta_i) \vert) . 
\label{center}
\eeq 
In addition, for each of the monopoles the phase associated with 
the unbroken U(1) is given by $p(\beta_i) /\vert p(\beta_i) \vert$.
The parameters in equation (\ref{alpbet}) are related to the parameters
in equation (\ref{pq}) by 

\beq 
\alpha_i={p(\beta_i)\over\prod_{j\not=i}(\beta_i-\beta_j)}, 
\eeq 
\beq 
b_j=(-1)^j\sum_{i_1<,\ldots,<i_j}\beta_{i_1},\cdots,\beta_{i_j}. 
\label{alphab}
\eeq 
Also if we define the vectors $A=({a_0,\ldots,a_{k-1}}),$
$P=(p(\beta_1),\ldots,p(\beta_k))$ then they satisfy the relation
\beq
A=BP
\label{abp}
\eeq 
where $B$ is the $k\times k$ Van der Monde matrix
\beq 
B=\left( \matrix{1 & \beta_1 & \beta_1^2 & \ldots \cr
1 & \beta_2 & \beta_2^2 & \ldots\cr
\vdots &&\vdots& \ddots} \right) \ .
\label{vdm}
\eeq
Solving for the $a_i$ parameters involves inverting the $B$ matrix and
is possible if and only if the $\beta_i's$ are distinct.

The remaining transformation to discuss is (1) we list above, namely 
the $SO(2)$ rotation about the preferred direction.  This may be 
found by transforming the parameters $a_i$ and $b_i$ as
\beq 
a_i\rightarrow\lambda^ia_i,\qquad b_i\rightarrow\lambda^{-i}b_i . 
\label{weight} 
\eeq 
It is useful to think of this transformation through an assignment 
of the various weights to the parameters under the rotation. 

Furthermore, a natural $2$-form defined on the space of the 
rational maps when all the coordinates $\beta_i$ and $p(\beta_i)$ 
are distinct is given by 
\beq 
\omega = \sum_{j=1}^k d\beta_i \wedge d\ln{p(\beta_i)}  \ .
\label{symplectic}
\eeq 
This form is closed and is symmetric under interchange of all $\beta_i$;
furthermore it is rational and defines a holomorphic symplectic form on 
$R_k$.  The existence of this $2$-form is directly related to the 
$2$-form $\omega_h$ made from the three K\"ahler forms in the twistor 
description of the moduli space.

The resultant is defined by 
\beq 
\Delta(p,q) \equiv \prod_{i=1}^{k} p(\beta_i)  \ ,
\eeq 
which is the denominator of the $2k$-form $\omega^k$.  
Physically, the center of the $k$-monopole is given in  
in eq.(\ref{center}) and the total phase is $\arg{\Delta(p,q)}$.  Setting 
$\Delta=1$ and $b_{k-1}=0$ corresponds to choosing the overall phase and 
location of the $k$-monopole to be zero.
The $SO(2)$ action in the $1-2$ plane acts on the resultant $\Delta$
trivially.  It is 
this action together with the weights given in eq.(\ref{weight}) 
we expect to maintain in taking the radius $R$ of the 
compactified $S^1$ away from zero.  

Note that the transformation (\ref{weight}) acts trivially on
the resultant.  One may assign a variant of this transformation
which acts nontrivially on the resultant \cite{segal}
\beq
a_i\rightarrow\lambda^{k-i}a_i,\qquad b_i\rightarrow\lambda^ib_i,\qquad
\Delta\rightarrow\lambda^{k^2}\Delta.
\eeq
It is this transformation which is more appropriate to gauge theories
and it will be related to the $U(1)_R$ symmetry.  There is another 
symmetry that the resultant respects. It is
given by a cyclic symmetry
\beq
a_i\rightarrow\omega a_i,\qquad \omega^k=1.
\label{cyclic}
\eeq 
Quotienting by this symmetry gives the proposed solution to the 
$N_f=0$ moduli space of the gauge theory.  

Let us compare the symmetries encoded in the rational map
to those in the three dimensional gauge theories.  The definition of 
a rational map
requires a preferred direction in $\RR^3$.  This space is identified
with the space in which the three scalars in the vector multiplet
transform as a triplet.  In section 2 we denoted
the double of the rotation group on this space $SU(2)_N$.
Choosing a preferred direction breaks this symmetry to $SO(2)_N$.  
The symmetry (1) is identified with the $SO(2)$ rotations about the
preferred direction parameterized by $\lambda$ in the last section.
The other symmetries parameterized by $\mu,\nu$ which correspond to
total phase and center of mass translations of the monopoles are
identified with overall scale tranformation, namely setting the 
dynamical scale $\Lambda$ to one and to the condition of
tracelessness on the $SU(N)$ matrices inside $U(N)$.

Let us move to compare the parameters in the rational map
to the asymptotic values of the gauge fields.
We recall that asymptotically for far apart monopoles, the positions of
the monopoles together with the phase of large gauge transformations
are given by 
\beq 
(\Re (\beta_i) ,\Im(\beta_i) , 
-(1/2k)\ln(\vert{p(\beta_i)}\vert,{p(\beta_i)\over
\vert{p(\beta_i)}\vert})
\eeq
where $\beta_i$ are defined in equation (\ref{alpbet}) and the polynomial
$p$ is defined in equation (\ref{pq}).
In view of the correspondence with gauge theories discussed
in section two, we identify these as
the classical expectation values of the scalar fields of the gauge
theory together with the dual photon
$(x^{(1)}_i,x^{(2)}_i,x^{(3)}_i,e^{i\theta_i})$, defined in equation
(\ref{phi}).
We see that the polynomial $q(z)$ is given asymptotically by the
characteristic polynomial of the scalar field $\phi$ on the $1-2$ plane
defind by $\phi=\phi_1+i\phi_2$,
$$q(z)\approx\det(z-\phi),$$
where $\phi_{1,2}$ are the scalar superpartners of the vector fields in
equation (\ref{phi}).
Singularities of this curve correspond to points in which at least
two monopoles coincide or, in the gauge theory language, an enhanced
$SU(2)$ group appears.
This lets us identify the coefficients $b_i$ in equation (\ref{pq}) 
asymptotically as the gauge invariant symmetric functions $s_i$ of the
scalar fields $\phi$, defined by Newton's formula
\beq
ks_k + \sum_{i=1}^k s_{k-i}tr(\phi^i) = 0 ,\qquad k=\{1,2,\ldots\} \comma
\label{Newton}
\eeq
where $s_0=1,s_1=tr(\phi)=0$.
The physical interpretation of the coefficients $a_i$ in equation (\ref{pq})
is slightly more complicated.  Using equations (\ref{abp}) and (\ref{vdm})
we find that they involve a symmetric combination of the locations on
the $1-2$ plane, the phases and locations in the 3rd direction. We will
give examples below. We may however see what the transformation
(\ref{cyclic}) means. It is equivalent to a constant shift in the dual
photons
\beq
\theta_i\rightarrow\theta_i+{2\pi j\over k},\qquad j=1,\ldots,k.
\eeq
Following this shift to the photons before dualizing it corresponds to
an action by the center of the gauge group $\ZZ_k$.  We learn that a quotient
by this symmetry for monopoles is equivalent to a quotient of the gauge
theory by its center.

\subsection{Example: $SU(2)$ gauge theory versus two monopoles} 

The rational map for a centered $2$ monopole and the resultant 
are given by
\beq
S(z)={a_0+a_1 z\over z^2+b_2},\qquad \Delta=a_0^2+b_2a_1^2=1.
\eeq
In \cite{ed} it was suggested that the resultant describes the
curve for $SU(2)$ gauge theory with one massless flavor.
The $\ZZ_2$ quotient (\ref{cyclic}) on the $a_i's$ is
$a_{0,1}\rightarrow-a_{0,1}$.  We can define invariants of this
quotient by $x=a_1^2, y=a_0a_1$ which results in a curve
$y^2+x^2b_2=x.$ This is the curve describing the Atiyah-Hitchin 
manifold for two centered monopoles or, as given by \cite{ed}, 
the curve for Yang-Mills gauge theory in three dimensions 
with $SU(2)$ gauge group.

We define for convenience $z_j=x^{(1)}_j+ix^{(2)}_j$,
$w_j=x^{(3)}_j+i\theta_j$. They satisfy the relations for
centered monopoles and zero phase $\sum_{i=1}^kz_i=\sum_{i=1}^kw_i=0$ 
which are the condition to choose an $SU(k)$ gauge group from $U(k)$.
Using equations (\ref{alphab}) - (\ref{vdm}) we find that 
approximately 
\beqar
a_0&\approx&\cosh w_1,\qquad
y\approx{\cosh w_1\sinh w_1\over z_1}\nonumber\cr
a_1&\approx&{\sinh w_1\over z_1},\qquad
x\approx{\sinh^2w_1\over z_1^2}\nonumber\cr
b_2&\approx&-z_1^2.
\eeqar
The dots denote higher inverse powers of $z_1$.  
Recently Bielawski \cite{bielawski} calculated the next order 
in the expansion around large expectation values $\beta_1,\beta_2$ for two
centered monopoles. The correction reads
\beq
|p(\beta_i)|^{1\over2}={1+\rho\over1-\rho}e^{x_i^3},\qquad
\rho={x^3_1-x^3_2\over|\vec x_1-\vec x_2|}
\eeq
Where the choice of coordinates is such that $\rho$ is positive.
This implies the form of correction from instanton contribution.
It would be interesting to verify that this is indeed correct.

The connection between the charge $2$ monopole and the $SU(2)$ 
gauge theory may be extended through the following.  In this case 
the rational map and resultant are obtained after a ``shift,''
\beq 
S(z)={(y+xz)\over (z^2-x^2+u)}, \quad\quad \Delta=y^2+x^2(u-x^2)  \ .
\label{twomon}
\eeq 
After the $Z_2$ quotient with the variables chosen in equation 
(\ref{twomon}) we obtain the elliptic curve describing the $N=2$ 
theory in four dimensions with gauge group $SU(2)$ (after 
including the point at infinity $x,y=\infty$).  We note by 
comparing to (\ref{twomon}) that
$u\approx z_1^2+{\sinh^2 w_1\over z_1^2}+\ldots$.  This allows 
us to compare to the $N=2$ $d=4$ relation between the 
period $a$ and the global coordinates $u$. 
The curve (\ref{twomon}) in the context of BPS monopoles describes 
two monopoles which are asymptotically located in the positions 
$\pm\sqrt{x^2-u}$ in the $x_1,x_2$ coordinates 
and in the $|y\pm x\sqrt{x^2-u}|$ in the $x_3$ direction. The argument 
of the last expression denotes the phases of the two monopoles.

A comment is in order on the scale dependence of the $d=4$ Yang-Mills theory.  
The resultant in general is free to equal any complex parameter other than 
zero; this corresponds to uncentering the monopole.  We recover the 
$\Lambda$ dependence in the (conventionally normalized) $SU(2)$ curve by 
taking its value to be  
\beq 
\Delta = {1\over 4}{\Lambda^4} \ ,
\eeq 
On the one hand, this identifies $\Lambda$ with the scale arising 
in the corresponding $d=3+1$ theory.  However, the center of mass 
of the monopoles is equal to $-1/4\ln(\vert\Delta\vert)$ 
while the total phase of the monopole is equal to $\arg(\Delta)$.  We 
can go over to the two centered monopole system with zero total 
phase by setting $\Delta=1$.  
We also have the freedom to make an overall scale transformation by 
setting $\Lambda=1$. This includes setting the vacuum angle
$\theta$ to zero. We learn that the center of mass motion of two 
monopoles is identified with a scale transformation and that changing 
the vacuum theta angle is identified with varying the total phase of
the two monopole system.  This identification naturally extends to all other
gauge groups.

\subsection{Example: $SU(3)$ and three monopoles}
For completeness we present a more general example, namely the 
curve proposed for the three-dimensional gauge group $SU(3)$.  
The rational map is
\beq
S(z)={a_0+a_1z+a_2z^2\over z^3+b_2z+b_3},
\eeq
The resultant is
\beq
\Delta=a_0^3+a_0a_1^2b_2-2a_0^2a_2b_2+a_0a_2^2b_2^2-a_1^3b_3
+3a_0a_1a_2b_3-a_1a_2^2b_2b_3+a_2^3b_3^2=1
\eeq
Invariants of the cyclic symmetry (\ref{cyclic}) are given 
by $A_1=a_0^2a_1, A_2=a_0a_1^2, A_3=a_0a_1a_2$.
After the quotient we obtain the result for the curve
\beq
A_0^3+A_0A_1^2b_2-2A_0^2A_2b_2+A_0A_2^2b_2^2-A_1^3b_3
+3A_0A_1A_2b_3-A_1A_2^2b_2b_3+A_2^3b_3^2=A_1A_2 \ .
\label{sutr}
\eeq
The $a_i$ parameters take the asymptotic form 
\beqar
a_0&\approx&{e^{w_1}z_2z_3\over(z_1-z_2)(z_1-z_3)}+{\rm perm.},\nonumber\cr
a_1&\approx&{e^{w_1}(z_2+z_3)\over(z_1-z_2)(z_1-z_3)}+{\rm perm.},\nonumber\cr
a_2&\approx&{e^{w_1}\over(z_1-z_2)(z_1-z_3)}+{\rm perm.} , 
\eeqar 
where the $z_i$ parameters are defined in the previous section.  
We propose that the equation (\ref{sutr}) describes the
moduli space of three dimensional $SU(3)$ Yang-Mills theory.

\subsection{BPS mass formula in three dimensions}

The symplectic form defined in equation (\ref{symplectic}) and
asymptotically in equation (\ref{forms}) serves as an integrand
for calculating masses of BPS saturated states.  Before stating
the formula a comment is in order. Equation (\ref{symplectic})
gives only two of the three real symplectic forms (the real 
and imaginary parts). This is
because in the rational maps description we are choosing a
preferred direction and the forms which it calculates in a
simple way are in the transverse plane to that direction.
The three K\"ahler forms sit in the triplet of $SO(3)$ 
and the third may be found by a rotation in $\RR^3$.

Recall that for the centered moduli space ${\cal M}_r^0$ of 
BPS monopoles in $SU(2)$, the fundamental group is 
$\pi_1({\cal M}_r^0)=Z^r$.  There are $r$ non-contractible cycles 
in which we may integrate a one-form around; this is seen 
in the asymptotic form of the manifold through the presence of a 
$T^r$.  A BPS mass formula for electrically charged 
states then reads
\beq
M=|\sum_in_i\vec a_i|,\qquad d\vec a_i=\oint_{\gamma_i}\vec\omega,
\label{BPS}
\eeq
where $n_i$ are charges with respect to the $U(1)$ gauge 
fields.  The set of $\gamma_i$ are a convenient choice of basis of 
one-cycles in the $H^1(M_r^0)$.  Note that the K\"ahler forms 
$\vec\omega$ and vectors ${\vec a}_i$ transform both as vectors 
under $SO(3)$.   

Magnetically charged states with respect to the $U(1)$
gauge fields are instantons in three dimensions. These are the fields
which contribute to the metric as discussed in section two.
Their mass formula will look similar to equation (\ref{BPS})
with electric charged being replaced by magnetic charges.

As an example we can calculate the semiclassical
mass of the $W_{ij}$ boson, given in equation (\ref{disc}), which
has electric charges $n_i=1$ of the gauge group $U(1)_i$ and
$n_j=-1$ of the gauge group $U(1)_j$ and $n_k=0$ in the rest.
We choose the basis of one cycles $\gamma_i$ to satisfy the
relation $\oint_{\gamma_i}d\tilde\theta_j=\delta_{ij}$.
Then we have asymptotically, using the approximate K\"ahler forms 
in equation (\ref{forms}), 
\beq
d{\vec a}_i=\oint_{\gamma_j}(d\tilde\theta_i\wedge d\vec x_i) , 
\label{vectors}
\eeq
where the integration over the first term in equation (\ref{forms})
vanishes. This calculation identifies asymptotically ${\vec a}_i$ 
with ${\vec x}_i$.  This occurs after matching the parameters $a_i$ 
and $x_i$ in the asymptotic regime and with the appropriate 
normalization in equation (\ref{vectors}).   
We obtain the expected relation to the mass formula, using (\ref{BPS}),
\beq 
M_{ij}=|\vec x_i-\vec x_j| . 
\eeq
Equation (\ref{BPS}) also serves as the BPS formula for a tension of
a string in four dimensions.
This follows from the correspondence between strings in four dimensions
and particles in three dimensions as discussed in the introduction.
When the BPS expression vanishes tensionless string arise in the 
spectrum.

\subsection{Comments on Spectral Curves}

The $R\rightarrow\infty$ limit in principle should reduce 
to the hyper-elliptic curves associated to $d=3+1$ $N=2$ super 
Yang-Mills theory, after including the appropriate points at 
infinity to obtain a compact surface.  There 
is another point of view associated with the equivalent spectral curve 
description of monopoles; we refer the reader to the 
references \cite{AH} for a complete description of the twistor 
construction.    

Associated to the charge $k$ monopole moduli space is a spectral curve in
$TCP_1$, the tangent bundle over $CP_1$, 
\beq
\eta^n + a_1(\xi) \eta^{n-1} + \ldots + a_n(\xi) = 0 \ .  
\label{spectral}
\eeq
The local coordinates in $TCP_1$ of the curve are given by $\rho {d\over
d\xi}$. For strongly centered monopoles the first coefficient $a_1(\xi)$ is
zero.

Holomorphic polynomials of order $2n$ over ${\rm CP}_1$ are defined by

\beq
a_n(\xi)= \sum_{j=0}^{2n} w_j \xi^j \ .
\label{etafunctions}
\eeq
The functions $a_n(\xi)$ are sections of line bundles ${\cal O}(2n)$ over
${\rm CP}_1$ consisting of all polynomials of order $2n$, and $\xi$ is the
${\rm CP}_1$ coordinate.  Furthermore, the standard representation of 
${\rm CP}_1$ is found by pasting together
two copies of the complex plane $\Phi$, $\tilde\Phi$ with coordinates $\xi$
and $\tilde\xi$; on the overlap $\Phi\cap{\tilde\Phi}$ the coordinates are
related by $\xi=1/{\tilde\xi}$.  On the overlap of the two
charts the functions in equation(\ref{etafunctions}) transform as ${\tilde
a}_n (1/\xi)= \xi^{-2n} a_n (\xi)$.

In this description a reality constraint on $a_n$ given by

\beq
{\bar a_n (\xi)} = (-1)^n {\bar\xi}^{2n} a_n
(-1/{\bar\xi}) \ ,
\label{reality}
\eeq
which means that $w_j = (-)^{n+j} {\bar w}_{2n-j}$. The operation in
equation (\ref{reality}) may be regarded as invariance under complex
conjugation together with the anti-podal map; in this case the real
structure is a map which takes $CP_1$ onto itself defined on the above
functions as $a_n (\xi) \rightarrow \bar a_n (-1/\bar\xi)$.

There is an additional map which acts on the ${\rm CP}_1$ coordinates
as $\xi \rightarrow -1/{\bar\xi}$ and $\eta\rightarrow -{{\bar \eta}\over 
{\bar \xi}^2}$. This is the analog of complex
conjugation on the twistor space and takes the complex structure on 
the twistor space to its inverse (i.e. complex conjugation 
$I\rightarrow -I$). Demanding
compatibility of the curve with the real structure enforces the 
coordinates $a_n(\xi)$ to satisfy the reality constraint in 
equation (\ref{reality}).

We now turn to the relation of monopoles moduli spaces to $N=4$ Yang-Mills
theories in three dimensions.  We proposed that $SU(n)$ Yang-Mills 
theories have moduli space of $n$ monopoles, obtained by 
quotienting the rational map by a cyclic group corresponding 
in the gauge theory description to the center.  
This result has a natural generalization to ADE groups which
is related to the discrete point groups of solids.
For $D_n$ series we propose that $SO(2n)$ Yang-Mills theories are 
obtained by quotienting the rational map of $2n$ monopoles by 
the dihedral group (generated by the cyclic generator and an 
inversion).  For the $E_n$ series a similar relation is proposed 
where the discrete group is appropriate to the $E_n$ group.

We would like to point out the following interesting procedure which
does not seem to agree with our proposal.
First recall the result of \cite{ed} for gauge group $SU(2)$.  
The moduli space for $N_f=0$ was identified as the Atiyah-Hitchin space. One
way to construct this space is to consider its simply connected double 
cover and then mod by the cyclic symmetry of
two monopoles - $Z_2$.
We will now describe a procedure due to Sutcliffe \cite{Sutcliffe} 
which relates the spectral curves to the curves associated with 
$d=4$ super Yang-Mills theory.
We recall the spectral curve for two monopoles
\beq
\eta^2 + a_1(\xi) \eta + a_2(\xi) = 0 
\eeq
The SO(3) spatial rotations of the 
underlying Euclidean space $R^3$ 
induce actions on the $TP_1$ coordinates (given for example in 
\cite{MM}).  The 
various quotients on the spectral curve we are describing translate 
directly into symmetric configurations of a multi-monopole.  
The $Z_2$ acts on the $TP_1$ coordinates as \beq
\eta\rightarrow-\eta\qquad\xi\rightarrow-\xi \label{symtwo}
\eeq
Imposing this symmetry and the reality condition equation
(\ref{reality}) we get a curve
for the $SU(2)$ gauge theory in four dimensions
\beq 
\eta^2+\xi^2u_2+\beta\xi^4+\bar\beta=0. 
\eeq
We mod by the symmetry (\ref{symtwo}) which has invariants 
$x={\eta\over\xi}$ and $z=\xi^2\beta$, set $\mu=|\beta|^2$ and get 
\beq 
z+{\mu\over z}+x^2+u=0. 
\eeq 
This is the curve 
for $SU(2)$ Yang-Mills theory as written in \cite{MW}.
The above is easily generalized to $SU(N)$ gauge theories and
we refer to \cite{Sutcliffe} for the details. 
Moreover we have performed a generalization of this to $D_n$ series.
Taking the spectral curve for $2n$ monopoles and moding by the symmetry
\beq
(\xi,\eta)\rightarrow(\omega\xi,\omega\eta),
\qquad(\xi,\eta)\rightarrow(-{\eta\over \xi^2},{1\over\xi}),
\eeq
We get the curve for $D_n$ series as written in \cite{MW}.
We expect the results to be generalized to $E_n$ groups in a
similar way.

The above construction is highly suggestive to make the 
connection with the hyper-elliptic curves of $N=2$ super 
Yang-Mills in four dimensions.  However, the coefficients in 
the curve are imposed to satisfy the reality constraint in 
eq.(\ref{reality}) which implies that the coefficients 
in the quotient of the curve are real; however, for gauge 
theories the parameters labeling the moduli space are 
complex.  The connection as a result is unclear.

\section{Discussion} 

The conjectured relation between $N=4$ low-energy effective actions 
and the moduli space of BPS solitons raises many interesting 
questions.   Besides expectations based on string theory, we would 
first like to point out that the asymptotic 
boundary conditions 
on the metric together with the appropriate $SO(3)$ action 
imposes a strong constraint on the correspondence between 
the gauge theory and monopole moduli spaces which needs to be 
investigated further.

One would like to verify that instanton corrections in the 
three dimensional theory actually reproduce the 
exponential corrections to the asymptotic form of the metric, 
which are alternatively known to arise in the monopole picture 
through exponentially damped charge exchange processes in the 
slow dynamics of magnetic monopoles.  For example, in $SU(2)$ 
with $N_f=0$ hypermultiplets these instanton corrections will 
exactly reproduce the Atiyah-Hitchin metric.  In this case, 
this statement is exact for the reason 
that the only $4$-dimensional hyper-K\"ahler metric 
in which one may write down that possesses the correct $SO(3)$ action 
gives rise to both 
the moduli space of $d=3$ vacua and the centered moduli space of 
a charge $2$ BPS monopole.  The correspondence 
for $SU(2)$ with $N_f=0$ predicts a non-trivial structure 
for the instanton corrections in the three-dimensional 
gauge theory; we suggest the same for higher-rank gauge 
groups.  

Further, whereas the dynamics of magnetic monopoles is modelled by 
a first quantized $N=4$ supersymmetric theory, the three 
dimensional gauge theory is a second-quantized 
field theory.  The connection between the two approaches is 
unresolved.  A possible explanation may arise as follows.  
All charge $k$-monopole moduli space metrics come 
from the inner product of the zero mode field deformations 
to the solutions of the BPS equations of a charge $k$ 
field solution in $SU(2)\rightarrow U(1)$.  From the $2+1$ dimensional 
gauge theory point of view the previous statement implies the 
existence of a master set of Bogolmony equations governing a space 
of theories with zero mode deformations responsible for the array of 
three dimensional nonperturbative $SU(N_c)$ $N=4$ 
super Yang-Mills theories.  The zero modes about the master 
field equation, acting on a functional with appropriate boundary 
conditions as in the monopole problem, would give rise to the 
gauge theories of different rank.  In this sense, an approach 
based on third quantization would explain how all the $SU(N_c)$ 
theories in $2+1$ dimensions with $N_f=0$ are related.
It would be interesting to explore this relation in more detail 
including its connection to the nonperturbative results of $N=2$ 
super Yang-Mills theories in four dimensions (i.e. taking the radius 
of the $S^1$ to infinity from the three-dimensional point of view).

The motivation to study three dimensional gauge theories came from
string theory as outlined in the introduction. Simple configurations
in string theory contain more matter multiplets in the fundamental
adjoint and various symmetric representations. For example a
configuration of intersecting 5-branes in $M$-theory will be associated
with the adjoint representation of $U(n)$ where $n$ is the charge of
one of the five branes. Such a gauge theory has a negative $s$ parameter
and therefore the semi-classical result is expected to be exact.
In particular no instanton like corrections are expected.  However
this suggests a ``one loop'' correction to the classical solution
in $M$-theory.  It would be interesting to check this.
Other applications of monopoles to string theory can be found by 
associating Manton scattering to five brane scattering in $M$-theory.

We conclude with several comments.  The appearance of closed geodesics 
appearing in the monopole moduli spaces should have an interesting 
analogy for the gauge theory.  The correspondence in the latter 
is plausibly related with the marginally unstable states occuring 
in the BPS spectrum.  

Finally we remark on a possible theory of ``nonabelian'' linear
multiplets which has part of its moduli space the monopole moduli space.
At points where the BPS mass formula vanishes we expect to get
tensionless strings in four dimensions.

\section{Acknowledgements} 

The work of G.C. and A.H. was supported in part by the NSF grant 
Nos. PHY-9309888 and PHY-9513835, respectively.  
We would like to thank J.~Distler, O.~Ganor, G.~Gibbons, 
K.~Intriligator, R.~Khuri, I.~Klebanov, N.~Seiberg for useful
conversations and especially E.~Witten for discussions.

\end{document}